\documentstyle[11pt,newpasp,twoside]{article}
\def\edcomment#1{\iffalse\marginpar{\raggedright\sl#1\/}\else\relax\fi}

\reversemarginpar

\begin{document}
\title{The MOJAVE Program: Studying the Relativistic Kinematics of AGN Jets}
\author{Matthew L. Lister}
\affil{Department of Physics, Purdue University, West Lafayette, IN 47907}

\begin{abstract}
We discuss a new VLBA program to investigate
structural and polarization changes in the brightest AGN
jets in the northern sky.  Our study represents a significant
improvement over previous surveys in terms of image fidelity, size,
and completeness, and will serve to characterize the kinematics of AGN
jets and determine how these are related to other source properties.
We discuss several preliminary results from our program, including the
detection of parsec-scale circularly polarized jet emission, enhanced
magnetic field ordering at the sites of apparent bends, and
intrinsic differences in the jets of strong- and weak-lined AGN.

\end{abstract}

\section{Introduction}

Kinematic studies using very-long-baseline interferometry have greatly
increased our understanding of extragalactic jets on parsec-scales,
but there are still many questions regarding their dynamics and
evolution that remain unanswered. This is partially due to the lack of
a long-term, systematic, full-Stokes imaging survey of a large
complete sample in which the selection effects are
well-understood. MOJAVE ({\bf M}onitoring {\bf O}f {\bf J}ets in {\bf
A}GN with {\bf V}LBA {\bf E}xperiments) is a new VLBA program that is
designed as a follow-up to the VLBA 2 cm survey (Kellermann et
al. 1998). It is currently investigating structural and polarization
changes in the 134 brightest AGN jets in the northern sky, of which 98
were part of the original VLBA 2 cm Survey. The main goals of MOJAVE
are to:

\begin{itemize}
\item{} Measure the distribution of apparent jet speeds, both within
individual sources and for the entire population.

\item{} Determine whether jet speed is related to other intrinsic
quantities, such as luminosity, black hole mass, and emission line
strength.

\item{} Characterize the kinematics of moving jet features, and
determine whether
they are consistent with streaming paths along bent trajectories. 

\item{} Investigate how magnetic fields evolve within
the jets, and how they track the moving features. 

\item{} Search for parsec-scale circularly polarized jet emission, and
examine how it evolves with time. 

\end{itemize}

\section{Sample Selection and Observational Status}
The goals of our program require a complete sample of extragalactic
jets that is large enough to investigate statistical aspects of the
parent population, and to perform inter-comparisons between various
sub-classes, such as quasars, BL Lacertae objects, gamma-ray loud
sources, and intra-day variables. Whereas the selection criteria of
the 2 cm VLBA survey were loosely defined in order to include a wide
range of sources such as gigahertz-peaked spectrum objects and radio
galaxies, the MOJAVE sample is selected purely on the basis of compact
radio flux density. The latter choice ensures a high degree of
completeness since a) all known AGN jets are radio-loud, b) all-sky
radio surveys are available at several frequencies, c) all our sources
are detectable by the VLBA. Another benefit is that we are able to
directly compare our data to Monte Carlo simulations of relativistic
beaming, without worrying about contamination from extended,
steep-spectrum emission. Also, the radio emission comes from the same
region as where we are measuring the apparent jet speeds, which is
important when estimating Doppler beaming factors from the kinematic
data. Our specific selection criteria are: a) declination $>
-20^\circ$, b) galactic latitude $|b| > 2.5^\circ$, and c) total 2 cm
VLBA flux density $\ge 1.5$ Jy at any epoch since 1994 ($\ge 2$ Jy for
sources with $\delta < 0^\circ$).

By not restricting our flux density criterion to a single epoch, we
have included many interesting variable sources that might have
otherwise been omitted.  Our final sample consists of 129 confirmed
and 5 candidate objects. We are currently gathering single-dish and
VLBA data on the latter to determine whether they meet our selection
criteria. Thirty-four of our sources are members of the third EGRET
gamma-ray catalog, and broken down by optical classification there are
95 quasars, 21 BL Lacs, 10 radio galaxies, and 8 unidentified
objects. Redshift information is currently available for $ 90\%$
of the sample.

Since the start of our observations in May 2002, we have obtained
single-epoch polarization images for 95\% of the sample, and 13
sources have been imaged at more than one epoch. As of Sept. 2003, we
have had ten successful observing sessions, of which nine have been
fully reduced. Our observations are currently planned to continue into
2004, and we have been allotted one new 24-hour long session every six
to eight weeks. We are able to observe eighteen sources each session,
which implies that each source is observed roughly once per
year. There are many highly variable sources that need more frequent
sampling, however, so we try to observe these more often at the
expense of longer monitoring intervals for the other sources. We are
also monitoring our sample at cm-wavelengths with the UMRAO and RATAN
telescopes to monitor overall spectral changes and to calibrate our
polarization vectors.

\section{Preliminary Results}

\subsection{Linear Polarization}

We have detected linearly polarized emission in all 123 jets we have
imaged thus far, with the exception of NGC 1052 and 2021+614. The
latter are likely heavily depolarized by intervening gas in the host
galaxy (see, e.g., Vermeulen et al. 2002).  The unresolved cores (located close to the base of the jet)
are weakly polarized, with the majority having fractional
polarizations under $ 4\%$. Six cores (3C 111, 4C 39.25, M87, 1413+135,
2008$-$159, and 2128$-$123) have no detectable polarization. In
most cases the magnetic field order increases with distance down the
jet, with features in the jets being appreciably more
polarized than the cores. In many sources (e.g., 1222+216;
Fig. 1), the field appears highly ordered on the outside edge where
the jet bends, suggesting a compression of the field from interaction
with the external medium.


\begin{figure}
\plotfiddle{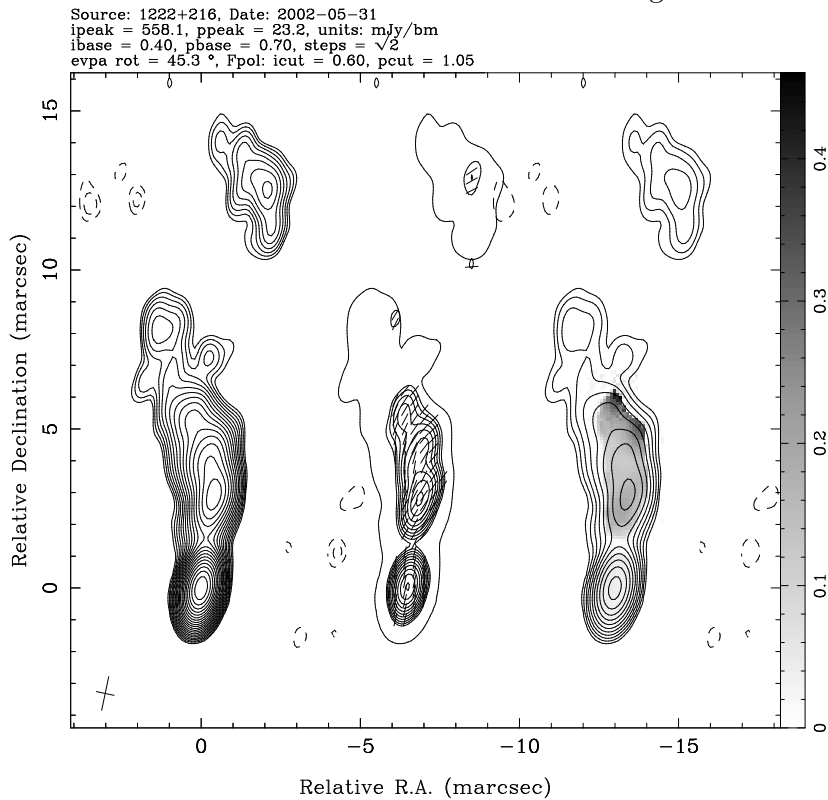}{3.8truein}{-90}{55}{55}{-215}{320}
\caption{\label{fracpol}\small VLBA 2 cm images of the quasar 1222+216.  Left image: total intensity contours. Middle
image: linear polarization contours with electric vectors
superimposed. Right image: total intensity contours with fractional
polarization greyscale ranging from 0 to 40\%. The restoring beam has
dimensions 1.08 $\times$ 0.57 mas, and the rms noise in the I image is
0.2 mJy/beam.}
\end{figure}

\subsection{Jet polarization versus optical line strength}

The question of whether the jets of weak-lined blazars (BL Lacertae
objects) are intrinsically different than those of broad-lined blazars
(radio-loud quasars) is still not fully resolved, in large part due to
uncertainties associated with projection and relativistic beaming
effects. One basic test involves looking for differences in
the jet polarization properties of these two classes. In
Figure~2 we plot the distribution of fractional
polarization for polarized features in the jets of BL Lacs and quasars
in our sample. A Kolmogorov-Smirnov test on the two distributions indicates
only a $0.1 \%$ probability that they are from the same parent
distribution.

\begin{figure}
\plotfiddle{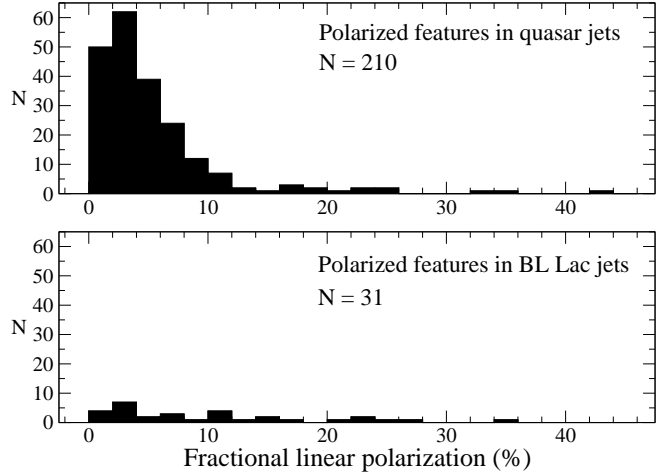}{2.5truein}{0}{37}{37}{-135}{-20}
\caption{\label{fracpol}\small Top panel: distribution of fractional linear polarization for
polarized features in quasar jets. Lower panel: distribution for
features in BL Lacertae jets.}
\end{figure}

Further evidence for intrinsic differences can be found in a plot of
fractional polarization versus distance from the core (Fig.~3). The
majority of the BL Lac jet components are more highly polarized {\it
at a given projected distance down the jet} than those of the
quasars. This confirms an earlier result obtained at 7 mm by Lister
(2001) for the smaller Pearson-Readhead AGN sample. Fig.~3 also shows
a weak trend of increasing fractional polarization down the jet, which
suggests that the magnetic fields of quasar jets take longer than the
BL Lac jets to become organized. It is noteworthy that these
polarization differences are located well outside the broad-line
region where the optical emission lines are produced. Further
investigation is required to determine whether these reflect intrinsic
differences in the interstellar medium of the host galaxy, or in the
jet itself.


\subsection{Circular Polarization}

The large number of sources that we observe at each epoch makes our
survey ideal for calibration of both linear and circular polarization
(CP). Using the techniques of Homan et al. (2001), we have detected weak
parsec-scale circularly polarized jet emission in 8 of 68 sources
analyzed thus far. Since the CP level is variable in many sources, we
anticipate on having a sufficient number of detections to look for
correlations between linear and circular polarization levels, and
monitor the CP sign consistency over time. We also intend to examine
whether circularly polarized jets possess any peculiar characteristics
that distinguish them from other sources.

\section{Summary}
The MOJAVE program represents the first large-scale, systematic survey
of a complete sample of AGN jets in all four Stokes parameters. Its
main goals are to obtain a better understanding of the kinematics and
magnetic field structures of relativistic jets, and determine how
these are related to other host galaxy properties such as the emission
line strength and black hole mass. A more thorough description of our
program and access to our on-line database can be found at
http://www.physics.purdue.edu/$\sim$mlister/MOJAVE.

The author wishes to acknowledge the members of the MOJAVE
collaboration: H. Aller, M. Aller, M. Cohen, D. Homan, M. Kadler,
K. Kellermann, Y. Kovalev, A. Lobanov, E. Ros, R. Vermeulen, and
J. Zensus.

\begin{figure}
\plotfiddle{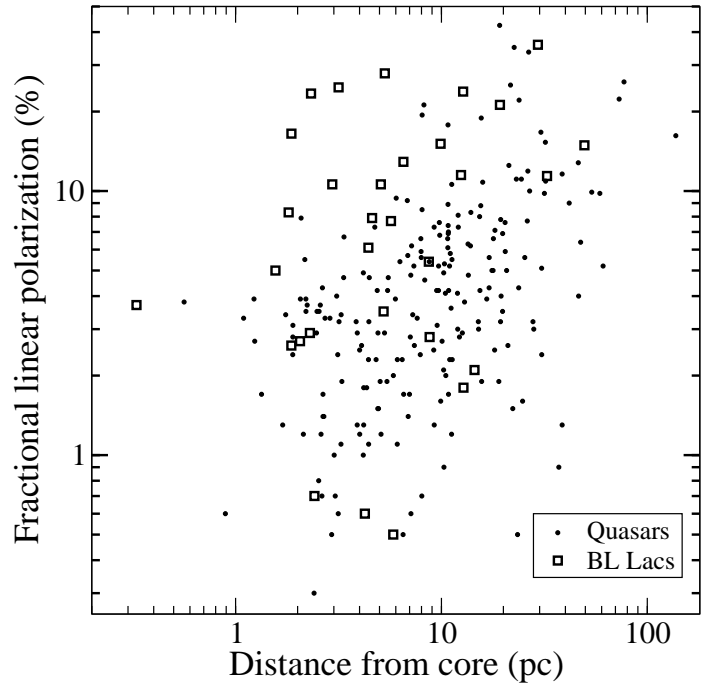}{3.7truein}{0}{50}{50}{-150}{-20}
\caption{\label{rvsmcpts}\small Plot of fractional linear polarization
versus projected distance from the core for
polarized features in the jets of quasars (circles) and BL Lacertae objects (open squares) .}

\end{figure}

\end{document}